\documentclass[10pt,conference]{IEEEtran}

\usepackage[english]{babel}
\usepackage{balance}
\usepackage{graphicx}
\usepackage{tikz}
\usepackage{multirow}
\usepackage{multicol}
\usepackage{booktabs}
\usepackage{url}
\usepackage{arydshln}
\usepackage{enumitem}
\usepackage{caption}

\usepackage[
    hypertexnames=false,
    colorlinks=true,
    citecolor=black,
    linkcolor=black,
    urlcolor=blue
]{hyperref}

\usepackage{cite}
\usepackage{listings}
\usepackage{mathtools}
\usepackage{amsmath}
\usepackage{xcolor}
\usepackage[normalem]{ulem}
\usepackage[english]{babel}
\usepackage[flushleft]{threeparttable}
\usepackage[all]{nowidow}
\usepackage{soul}
\usepackage{ListStyle}
\usepackage{bm}
\usepackage{tablefootnote}
\usepackage{threeparttable}
\usepackage{array, makecell}
\usepackage{colortbl}
\usepackage{amsfonts}
\usepackage{float}
\usepackage{wrapfig}
\usepackage{pifont}  
\usepackage{tcolorbox}
\usepackage{textcomp}
\usepackage{placeins}

\definecolor{newGreen}{rgb}{0.0,0.63,.333}
\definecolor{newYellow}{rgb}{1.0,.7,.1}

\useunder{\uline}{\ulined}{}%
\DeclareUrlCommand{\bulurl}{}

\newcommand{\mytool}{{NumAnchor}}

\def\halfcheck{\textcolor{newYellow}{\checkmark\kern-1.1ex\raisebox{.7ex}{\rotatebox[origin=c]{125}{--}}}}

\colorlet{tableheadcolor}{gray!25} 
\colorlet{tablerowcolor}{gray!10} 
\title{Beyond Tokens: NumAnchor for Address-Grounded Numeric Operand Representation in Assembly Language Models}

\author{
\IEEEauthorblockN{
Kun Liu\IEEEauthorrefmark{1},
Keith Mitchell\IEEEauthorrefmark{1},
Xin Hu\IEEEauthorrefmark{1},
Jiangcheng Yu\IEEEauthorrefmark{3},\\
Chenke Luo\IEEEauthorrefmark{1},
Zhengming Ding\IEEEauthorrefmark{1},
Haotian Zhang\IEEEauthorrefmark{2},
Jiang Ming\IEEEauthorrefmark{1}
}
\IEEEauthorblockA{
\IEEEauthorrefmark{1}Tulane University, USA\\
\IEEEauthorrefmark{2}New Jersey Institute of Technology, USA\\
\IEEEauthorrefmark{3}Independent Researcher
}
}

\begin{document}

\maketitle

\pagestyle{empty}
\thispagestyle{empty}

\begin{abstract}
Assembly language models (ALMs) are becoming important foundations for automated binary program analysis, but their token-level interfaces remain mismatched with numeric-like operands, which are common and semantically dense in stripped binaries. In two evaluated corpora, such operands account for 45.7\% and 40.9\% of all operand occurrences, yet their surface values may denote branch targets, global objects, stack slots, ordinary constants, or unresolved pointer-derived values. Existing ALMs expose this heterogeneity only partially through coarse placeholders, jump-specific rewriting, or execution-derived side channels, leaving a representation-boundary gap between static-analysis preprocessing and model-facing inputs.

We present \textit{NumAnchor}, a structure-aware input-boundary adapter that exposes recoverable numeric-like operands as typed, canonicalized program entities for ALMs. Rather than proposing a new static analyzer or modifying Transformer backbones, NumAnchor defines how recovered code, data, and stack artifacts enter the model interface: it preserves repeated entity identity and locality through compact side descriptors and fuses them before the first Transformer layer while keeping host architectures, pretraining objectives, and downstream protocols unchanged.  
On frozen instruction-level embeddings, NumAnchor makes address-grounded structure recoverable, reaching 98.7\% hierarchical-location probing accuracy and 0.932/0.907 AUROC on jump/call target alignment. Across instruction-level, function-level, rebasing-oriented, and execution-aware ALMs, it improves 10k-candidate function retrieval by up to 35.0\% in Recall@1 and 31.7\% in MRR, and improves execution-aware type recovery by up to 5.2 F1 points.
Ablations show complementary gains from code, data, and stack descriptors, while the side-channel design keeps overhead modest without increasing host sequence length.
\end{abstract}


\section{Introduction}
\label{sec:Introduction}


Deep learning has become a central paradigm for automated software engineering and program analysis~\cite{allamanis2018learning,alon2019code2vec,le2020deep}. In binary program analysis, models often operate on executable artifacts rather than source code, where semantics are reflected through low-level representations and compiler-shaped structures~\cite{angr2016artofwar}. Some approaches lift binaries into higher-level representations to reuse source-oriented models~\cite{chen2022,pal24sp,ase19lacomis}, but such lifting can obscure executable-level layout and address-grounded structure. Assembly language models (ALMs) offer a complementary path by learning directly from disassembly. Pre-trained on large unlabeled assembly corpora, ALMs provide transferable representations for binary-analysis and reverse-engineering tasks, including function recognition~\cite{Shin2015}, type recovery~\cite{Chua2017}, binary similarity~\cite{ding2019asm2vec}, vulnerability search~\cite{Luo2023VulHawkCV,Arasteh24,arasteh2025}, and malware analysis~\cite{Downing2021,chen2023,Pack-ALM}.


Yet, learning directly from disassembly does not remove the representation mismatch between binary code and text-based modeling~\cite{heusenix24}. Most ALMs follow a Transformer-based sequence-modeling paradigm, operating on tokenized instruction streams rather than explicitly representing the structural relations that organize binary programs~\cite{vaswani2017attention}. Although disassembly is serialized as mnemonics and operands, many assembly tokens carry recoverable program structure whose meaning cannot be inferred from surface form alone. This issue is especially prominent for numeric-like operands, which account for \textbf{45.7\%} and \textbf{40.9\%} of all operand occurrences in the StateFormer~\cite{pei2021stateformer} and jTrans~\cite{wang2022jtrans} corpora, respectively. In a stripped binary, a hexadecimal value such as \texttt{0x3ccd4} in \texttt{jmp 0x3ccd4} is not merely a constant; it denotes a control-flow anchor. Raw numeric values are also unstable across relocation, rebasing, and compiler transformations. Treating them as ordinary lexical tokens encourages models to learn brittle layout-specific artifacts, whereas uniformly normalizing them into generic placeholders discards entity identity and locality needed by downstream binary-analysis tasks.

Prior work has recognized the difficulty of modeling numeric operands, but existing interfaces expose recovered structure only partially. Abstraction-based ALMs~\cite{li2021palmtree,koo2021deepsemantic,binshot} reduce out-of-vocabulary (OOV) effects by replacing numerics with coarse placeholders, often losing repeated entity identity and locality. Jump-aware methods~\cite{wang2022jtrans,wang2024clap} preserve selected control-flow targets, but provide limited support for data references and stack layouts. Execution-aware models~\cite{pei2021stateformer,pei2020trex,pei2022neudep} expose runtime semantics through traces or state sequences, yet these signals may be unavailable in static deployment. Thus, the missing piece is not merely more static-analysis artifacts, but a unified model-facing interface that converts recoverable code, data, and stack structure into typed, layout-stable signals for ALMs.


To address this representation-boundary gap, we present \emph{\mytool}, a structure-aware input-boundary interface for numeric-like operands in ALMs. Rather than introducing a new static-analysis algorithm or relying on task-specific rewriting rules, \mytool\ defines how recovered static artifacts enter the model interface. It maps recoverable numeric-like operands to typed and canonicalized program entities, preserves shared identities and hierarchical locality through compact side descriptors, and falls back to generic numeric representations when grounding is unavailable or ambiguous. These signals are fused at the input embedding layer, allowing \mytool\ to integrate with existing Transformer-based ALMs without changing their backbone architectures, pretraining objectives, or downstream protocols.


We evaluate \mytool\ as a non-invasive adapter across multiple binary-analysis tasks and ALM backbones. Frozen-embedding probes show that \mytool\ makes address-grounded structure recoverable after pretraining, reaching 98.7\% hierarchical-location accuracy and 0.932/0.907 AUROC on jump/call target alignment. For function retrieval, \mytool\ improves instruction-level and function-level ALMs by up to 35.0\% in Recall@1 and 31.7\% in MRR in the challenging 10k-candidate setting. For type recovery, it improves execution-aware modeling by up to 5.2 F1 points. Ablations show that code, data, and stack descriptors provide complementary benefits over role-only and generic numeric encodings, with modest parameter and inference overhead.

In summary, this paper makes the following contributions:

\begin{itemize}
\item \textbf{Problem Characterization.}
We identify numeric-like operands as a central representation challenge for ALMs in stripped binaries. We show that existing tokenization and normalization strategies struggle to preserve transferable structural meaning from raw numeric values.

\item \textbf{Structure-Aware Representation Interface.}
We introduce \emph{\mytool}, a representation-boundary interface that exposes recoverable program structure to ALMs in a stable and model-compatible form. Rather than proposing a new static analyzer or task-specific heuristics, {\mytool} provides a unified mechanism for preserving structural signals before binary-code tokenization.

\item \textbf{Evaluation Across Downstream Tasks.}
We evaluate \mytool{} across representative ALM families, input granularities, and downstream binary-analysis tasks. The results show consistent gains over baseline models and simpler numeric encodings, while ablations confirm the complementary value of different structural signals with modest overhead.
\end{itemize}

\section{Background and Related Work}
\label{sec:Background}

\begin{figure}[t]
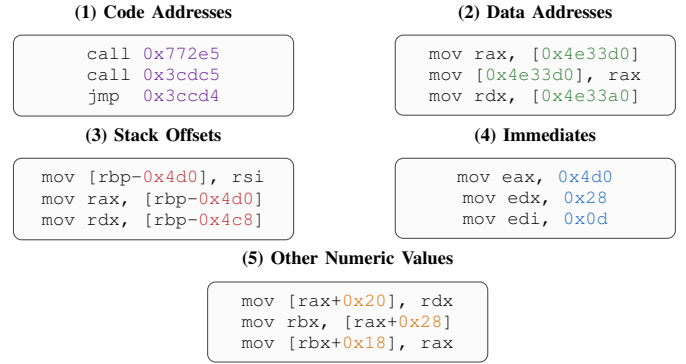

\centering
\scriptsize

\newcommand{\operandbox}[2]{
\begin{minipage}[t]{0.42\linewidth}
\centering
\textbf{#1}\\[-0.8mm]
\begin{tcolorbox}[
    colback=gray!3,
    colframe=black!70,
    boxrule=0.4pt,
    arc=1mm,
    width=\linewidth,
    left=1mm,
    right=1mm,
    top=0.5mm,
    bottom=0.5mm
]
\centering
\ttfamily
#2
\end{tcolorbox}
\end{minipage}
}

\operandbox{(1) Code Addresses}{
call \ctgt{0x772e5}\\
call \ctgt{0x3cdc5}\\
jmp \ \ctgt{0x3ccd4}
}
\hfill
\operandbox{(2) Data Addresses}{
mov rax, [\daddr{0x4e33d0}]\\
mov [\daddr{0x4e33d0}], rax\\
mov rdx, [\daddr{0x4e33a0}]
}

\vspace{1.5mm}

\operandbox{(3) Stack Offsets}{
mov [rbp-\stk{0x4d0}], rsi\\
mov rax, [rbp-\stk{0x4d0}]\\
mov rdx, [rbp-\stk{0x4c8}]
}
\hfill
\operandbox{(4) Immediates}{
mov eax, \immv{0x4d0}\\
mov edx, \immv{0x28}\\
mov edi, \immv{0x0d}
}

\vspace{1.5mm}

\begin{minipage}{0.42\linewidth}
\centering
\textbf{(5) Other Numeric Values}\\[-0.8mm]
\begin{tcolorbox}[
    colback=gray!3,
    colframe=black!70,
    boxrule=0.4pt,
    arc=1mm,
    width=\linewidth,
    left=1mm,
    right=1mm,
    top=0.5mm,
    bottom=0.5mm
]
\centering
\ttfamily
mov [rax+\unk{0x20}], rdx\\
mov rbx, [rax+\unk{0x28}]\\
mov [rbx+\unk{0x18}], rax
\end{tcolorbox}
\end{minipage}

\caption{Representative numeric-like operands from \texttt{openssl}'s main function.}
\label{fig:openssl_numeric_operand_examples}
\vspace{-6mm}
\end{figure}


This section explains why numeric-like operands in stripped binaries are difficult to model with standard token interfaces, and then reviews how prior ALMs expose such operands after preprocessing. This discussion identifies the representation-boundary gap that motivates \mytool{}.

\begin{table*}[t]
\centering
\small
\setlength{\tabcolsep}{4pt}
\renewcommand{\arraystretch}{1.15}
\captionsetup{skip=2pt}
\begin{threeparttable}
\caption{Address/entity signals explicitly exposed for numeric-like operands at the model interface after preprocessing. $\checkmark$: explicit support; $\triangle$: partial, indirect, or case-limited support; $\times$: no explicit support; $^\dagger$: used only during pretraining or through an auxiliary channel.}

\label{tab:address_capability}
\begin{tabular}{lcccccc}
\toprule
\textbf{Method}
& \textbf{Entity}
& \textbf{Same}
& \textbf{Branch}
& \textbf{Internal}
& \textbf{Instruction}
& \textbf{Structural} \\
& \textbf{role}
& \textbf{entity}
& \textbf{target}
& \textbf{callee}
& \textbf{location}
& \textbf{locality} \\
\midrule

PalmTree
& $\triangle$ & $\times$ & $\times$ & $\times$ & $\times$ & $\times$ \\

DeepSemantic / BinShot / UniASM / BinBert
& $\triangle$ & $\times$ & $\triangle$ & $\triangle$ & $\times$ & $\times$ \\

jTrans
& $\triangle$ & $\times$ & $\checkmark$ & $\times$ & $\times$ & $\triangle$ \\

CLAP
& $\triangle$ & $\times$ & $\checkmark$ & $\triangle$ & $\times$ & $\triangle$ \\

TREX / StateFormer
& $\times$ & $\triangle^\dagger$ & $\times$ & $\times$ & $\times$ & $\times$ \\

NeuDep
& $\times$ & $\triangle^\dagger$ & $\times$ & $\times$ & $\triangle^\dagger$ & $\triangle^\dagger$ \\

\midrule
\textbf{Ours}
& $\checkmark$ & $\checkmark$ & $\checkmark$ & $\checkmark$ & $\checkmark$ & $\checkmark$ \\

\bottomrule
\end{tabular}


\begin{tablenotes}[para,flushleft]
\footnotesize
\item[] \textit{Note.} The table compares how recovered artifacts are exposed at the model interface after preprocessing, rather than overall model performance. \textit{Entity role} denotes explicit typing of numeric operands as code-, data-, stack-, or value-like entities. \textit{Same entity} denotes stable co-reference for repeated recovered entities. \textit{Branch target} and \textit{Internal callee} denote preserved direct branch destinations and internal call targets. \textit{Instruction location} denotes exposure of the current instruction's recovered code anchor. \textit{Structural locality} denotes retained normalized layout information across function/block, section, or stack-frame structure.
\end{tablenotes}

\end{threeparttable}
\vspace{-5mm}
\end{table*}

\subsection{Why Numeric-like Operands Are Hard to Model}
\label{sec:bg_numeric}

\noindent
\textbf{\textit{Large and Unstable Value Space.}}
Modeling numeric-like operands in stripped binaries creates a mismatch with standard text-based representations. Transformer-based ALMs rely on finite token vocabularies, whereas numeric operands may range over large $32$-bit or $64$-bit address spaces. Treating each raw literal as an independent token is therefore impractical: many values are rare or unseen during training, and their embeddings provide little reusable signal across binaries. Moreover, raw numeric values often change across compiler optimizations, rebasing, relocation, and memory-layout changes, encouraging models to memorize layout-specific artifacts rather than transferable program structure.

\vspace{2pt}
\noindent
\textbf{\textit{Position-Dependent Semantics.}}
The meaning of a numeric operand depends on where and how it appears in the executable, not on its surface value alone. As illustrated in Figure~\ref{fig:openssl_numeric_operand_examples}, the literal \texttt{0x4d0} may represent a base-relative stack-frame offset (Box 3) or an ordinary arithmetic immediate (Box 4). Similarly, hexadecimal-looking operands may denote control-flow targets (Box 1), global data objects (Box 2), stack slots (Box 3), or unresolved pointer-derived displacements (Box 5). Thus, numeric operands cannot be interpreted reliably from lexical form alone; they must be grounded in their surrounding program structure.

\vspace{2pt}
\noindent
\textbf{\textit{Raw Values vs. Over-Abstraction.}}
These properties create a practical tension for ALM input design. Preserving raw literals keeps some structural distinctions visible, such as whether two instructions refer to the same global object or stack slot, but exposes the model to unstable layout-specific values. Conversely, normalizing all numerics into coarse placeholders, such as \texttt{<IMM>} or \texttt{<ADDR>}, reduces vocabulary sparsity but removes the structural distinctions that many binary-analysis tasks require. A useful representation should therefore balance these extremes: it should preserve grounded structural identities when static analysis is confident, distinguish roles such as code references, data references, stack-relative slots, and ordinary immediates, and fall back to generic numeric representations when grounding is ambiguous.

\subsection{How Existing ALMs Handle Numeric-like Operands}
\label{sec:bg_prior}

The challenges discussed in Section~\ref{sec:bg_numeric} provide a useful lens for examining how existing ALMs represent numeric-like operands. As summarized in Table~\ref{tab:address_capability}, prior work exposes different subsets of entity-level information, but existing interfaces do not uniformly preserve layout-stable structure across code, data, and stack operands.

\vspace{2pt}
\noindent
\textbf{\textit{Lexical Abstraction and Role-Aware Normalization.}}

To reduce vocabulary sparsity and out-of-vocabulary (OOV) cases, representative ALMs, including PalmTree~\cite{li2021palmtree}, DeepSemantic~\cite{koo2021deepsemantic}, BinShot~\cite{binshot}, UniASM~\cite{uniasm}, and BinBert~\cite{binbert}, apply instruction-level or role-aware normalization to numeric operands. These approaches replace raw numeric literals with symbolic or categorical placeholders, such as \texttt{<ADDR>}, \texttt{<IMM>}, or other role-specific variants. This prevents models from directly memorizing volatile surface values and exposes basic operand categories. However, because many distinct entities are still collapsed into shared placeholders, this representation does not preserve fine-grained entity identity or stable locality, such as whether distant instructions access the same global object or stack slot.

\vspace{2pt}
\noindent
\textbf{\textit{Jump-Aware Symbolic Rewriting.}}
A second line of work incorporates limited control-flow structure into the input representation. For example, jTrans~\cite{wang2022jtrans} models jump relationships by modifying positional embeddings and rewriting direct jump targets as symbolic labels with relative destination offsets. CLAP~\cite{wang2024clap} instead combines deterministic address rebasing with subword WordPiece tokenization~\cite{kudo2018wordpiece}. These designs help preserve local branch-target relationships and expose control-flow information to the host model. However, they are mainly specialized for jump-aware control-flow targets. They provide limited support for data references and stack-relative layouts, and sequence-relative indicators can still be constrained by the host Transformer's context window, especially for long-range intra-procedural jumps and inter-procedural \texttt{call} targets.

\begin{figure*} [t]
	\centering
	\includegraphics[width=0.98\textwidth,angle=0,scale=1.0]{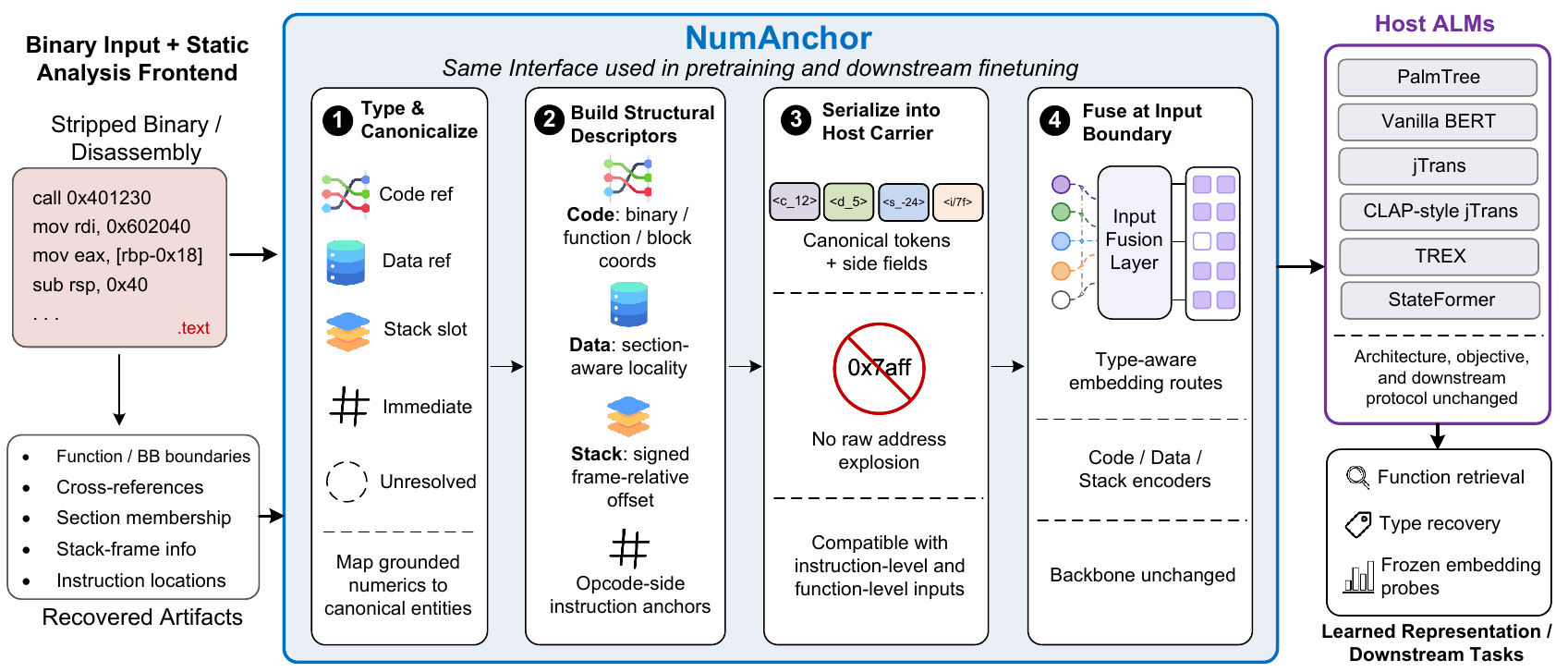}
	
\caption{Overview of \mytool: a representation-boundary interface that exposes static-analysis-recovered numeric-like operands and instruction locations as typed, structure-aware signals for host ALMs.}	
  \label{fig:overview}
	\vspace{-4mm}
\end{figure*}

\vspace{2pt}
\noindent
\textbf{\textit{Execution-Aware Dynamic Channels.}}
A third class of ALMs, including TREX~\cite{pei2020trex}, StateFormer~\cite{pei2021stateformer}, and NeuDep~\cite{pei2022neudep}, augments the representation with dynamic execution traces, register states, or memory states. To process these numerical fields and temporal execution patterns, these pipelines often introduce specialized continuous or sequential components, such as NAU~\cite{madsen2020neural} or BiLSTM encoders~\cite{Graves2005602}. This paradigm is effective for exposing operational semantics during pre-training, including runtime memory dependencies, instruction effects, and chronological value transformations. 
However, such auxiliary signals may be unavailable in static downstream settings.

\vspace{2pt}
\noindent
\textbf{\textit{Remaining Gap.}}
Overall, existing strategies address important parts of numeric operand modeling, but they do not provide a unified, layout-normalized interface for static binary-analysis tasks. Lexical abstraction controls vocabulary size but can collapse distinct roles and repeated entities; for example, \texttt{0x4d0} in \texttt{[rbp-0x4d0]} denotes a stack-frame slot, whereas \texttt{0x4d0} in \texttt{mov eax, 0x4d0} is an ordinary immediate. Jump-aware rewriting captures selected branch targets, such as \texttt{jmp 0x3ccd4}, but does not uniformly expose internal call targets such as \texttt{call 0x772e5}, nor does it generalize to data and stack structure, such as repeated references to the global object \texttt{[0x4e33d0]}. Execution-aware channels can expose runtime behavior for pointer-derived offsets, such as \texttt{[rax+0x20]}, but rely on signals that may be unavailable during static deployment. Moreover, components such as NAU and BiLSTM encoders are better suited for continuous magnitudes and sequential dependencies than for representing binary addresses as stable topological anchors. These limitations suggest that the key gap is not the absence of static-analysis artifacts, but the lack of a unified model-facing interface for exposing numeric-related structure to ALMs.

\section{Motivation and Overview}
\label{sec:overview}

The limitations discussed in Section~\ref{sec:Background} point to a representation-boundary gap between static-analysis preprocessing and ALM input design, rather than a lack of recoverable program structure. Static-analysis artifacts are already a routine part of binary-analysis and ALM preprocessing pipelines, where recovered information is commonly used to construct or refine model inputs~\cite{pei2021xda,yu2022deepdi}. 
For numeric-like operands, however, these artifacts are often consumed only as preprocessing aids: they help normalize, filter, or serialize token streams, but their entity roles, co-reference, and locality are not exposed to the model as unified, learnable signals. As a result, existing ALMs either collapse numeric operands into coarse placeholders, expose only narrow task-specific cues, or rely on execution-derived signals that may be unavailable in static deployment.


To address this gap, we present \emph{\mytool}, a backbone-agnostic adapter that operates at the input boundary. Rather than introducing a new static-analysis algorithm or modifying the Transformer backbone, \mytool\ defines how recovered artifacts enter the model interface. It maps recoverable numeric-like operands to typed and canonicalized program entities, attaches compact descriptors for shared identity and hierarchical locality, and falls back to generic numeric representations when static grounding is unavailable or ambiguous. These descriptors turn brittle numeric literals into reusable, layout-stable signals for downstream ALMs.

\subsection{\mytool\ Workflow}
\label{sec:workflow}

Figure~\ref{fig:overview} shows \mytool{} as a representation-boundary interface between a static-analysis frontend and the host ALM. Given raw disassembly and recovered structural artifacts, \mytool{} converts numeric-like operands into typed, canonicalized, and structurally augmented input signals. The same interface is used during pre-training and downstream fine-tuning, ensuring that learned structural cues remain available during static deployment.

\vspace*{2pt}
\noindent\textbf{\textit{I. Type and Canonicalize Entities ($\S\ref{sec:design_preprocessing}$).}}
\mytool{} classifies numeric-like operands into code, data, stack, immediate, or unresolved categories, and canonicalizes grounded operands into shared model-facing entities.

\vspace*{2pt}
\noindent\textbf{\textit{II. Construct Structural Descriptors ($\S\ref{sec:design_encoding}$).}}
For grounded entities, \mytool{} builds compact descriptors that capture code locality, data-section layout, and stack-frame structure in a normalized or relative form.

\vspace*{2pt}
\noindent\textbf{\textit{III. Serialize into the Host Carrier ($\S\ref{sec:design_serialization}$).}}
\mytool{} serializes canonical tokens and structural side fields into the host model's native input format without expanding raw address tokens.

\vspace*{2pt}
\noindent\textbf{\textit{IV. Fuse at the Input Boundary ($\S\ref{sec:design_fusion}$).}}
Finally, \mytool{} injects these signals before the first Transformer layer, preserving the host backbone, objectives, and downstream protocols.

\begin{figure*} [t]
	\centering
	\includegraphics[width=0.95\textwidth,angle=0,scale=1.0]{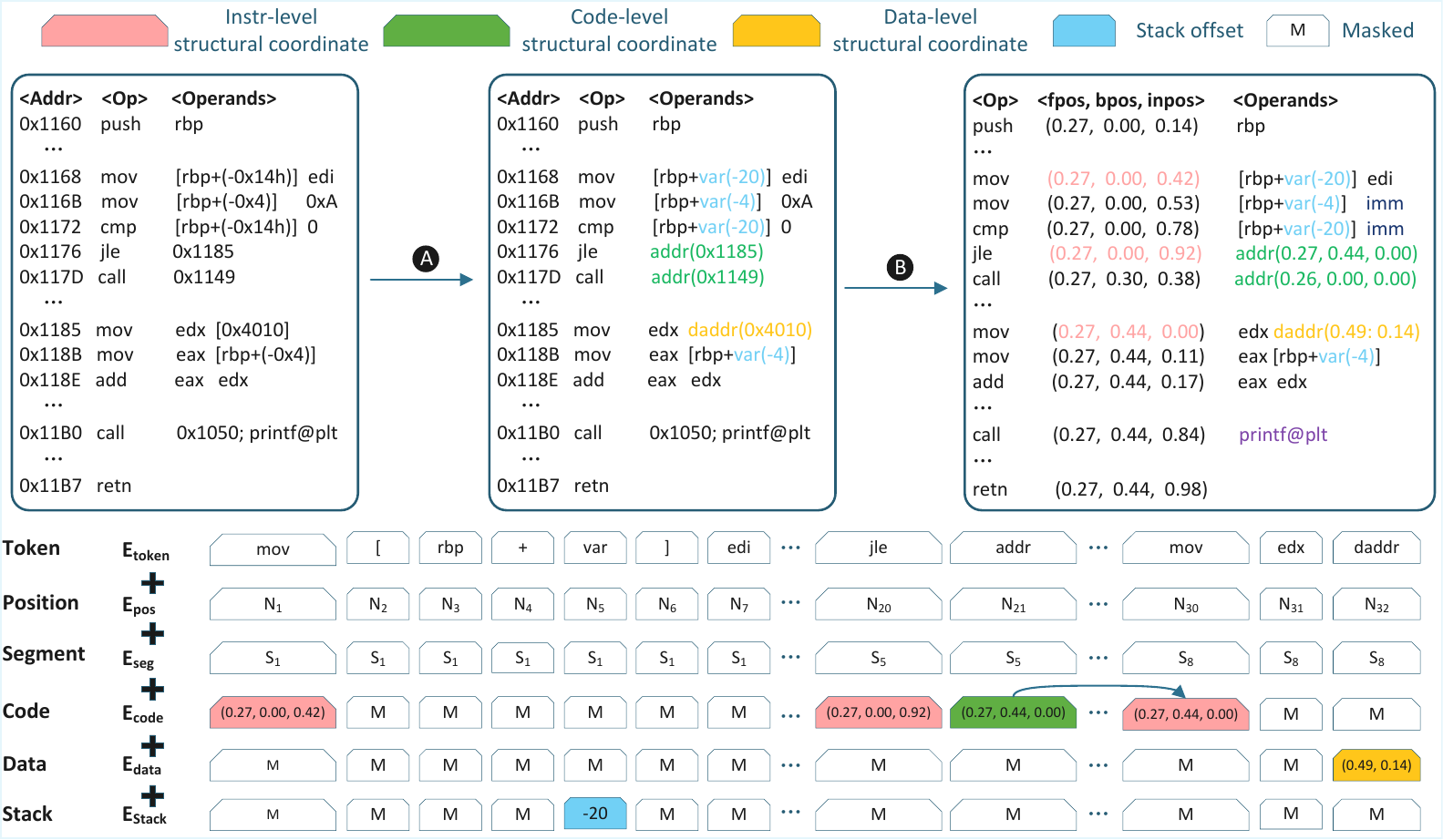}
    \caption{NumAnchor at the token boundary. Step~A canonicalizes raw numeric-like operands into compact typed tokens while preserving resolved entities internally. Step~B attaches route-specific side fields for code, data, and stack structure; masked entries denote inactive routes. Structural coordinates are rounded to two decimal places for readability.}
	\label{fig:systemdesign}
\end{figure*}

\section{System Design}
\label{sec:systemdesign}


Figure~\ref{fig:systemdesign} summarizes \mytool{}'s token-boundary workflow. \mytool{} operates as an input-boundary adapter: it leaves the host Transformer unchanged while augmenting the representation that enters it. Given disassembly and recovered static-analysis artifacts, \mytool{} follows the four-stage pipeline introduced in Figure~\ref{fig:overview}: (i) it types numeric-like operands and opcode-side instruction anchors, then canonicalizes them into compact model-facing tokens; (ii) it constructs type-specific side descriptors; (iii) it serializes these descriptors into the host carrier; and (iv) it fuses them before the first Transformer layer.

For any disassembled unit, NumAnchor emits a host-compatible sequence
\[
X = \big[(x_i,r_i,p_i,o_i,z_i)\big]_{i=1}^{L},
\]
where $x_i$ is the lexical token processed by the host model, $r_i$ is the active NumAnchor routing path, $p_i$ is an optional multi-dimensional code- or data-structure descriptor, $o_i$ is an optional signed stack offset, and $z_i$ encapsulates host-specific auxiliary fields. The semantic routing space is defined as
\[
\mathcal{R}=\{\mathrm{code},\mathrm{data},\mathrm{stack},\mathrm{imm},\mathrm{num},\bot\},
\]
where $\bot$ denotes tokens with no NumAnchor-specific route. During preprocessing, NumAnchor also maintains an internal grounded entity identifier $e_i$ for recovered code and data entities to resolve co-reference. This identifier is used only to construct structural descriptors and is never emitted as a standalone lexical token, avoiding expansion of the host token sequence. As a result, NumAnchor preserves the host backbone architecture, pre-training objectives, and downstream execution protocols.

\subsection{Analysis-Driven Typing and Canonicalization}
\label{sec:design_preprocessing}

\noindent
\textbf{\textit{Typing}.}
Typing systematically maps each numeric-like operand to a specific semantic routing path based on recovered static-analysis artifacts, such as cross-references, section membership, function/basic-block boundaries, and stack-frame layouts. Our prototype extracts these artifacts using IDA Pro~\cite{hexrays_idapro}, but the interface can be instantiated with other disassembly frontends that provide comparable artifacts. For each operand occurrence $v_i$ within an analysis context $c_i$, typing
assigns a routing path
\[
r_i = T(v_i,c_i),
\]
where $T$ maps operands to one of the \mytool{} semantic routes defined above. These routes differentiate internal code references, static data references, stack slots, ordinary constants, and unresolved values. Additionally, opcode tokens retain their original lexical forms but are bound to the current instruction address as an optional code anchor when localized; all other tokens are forwarded with $r_i=\bot$.

\vspace{2pt}
\noindent
\textbf{\textit{Canonicalization}.}
Canonicalization decouples semantic identities from surface literals by mapping a typed occurrence to a compact lexical token together with internal grounding metadata:
\[
\Gamma(v_i,c_i,r_i) \mapsto (x_i,e_i,o_i).
\]
Here $x_i$ denotes the emitted model-facing token, $e_i$ denotes an internal grounded identifier for resolved code, data, or instruction-anchor entities, and $o_i$ records an optional signed stack displacement. 
The identifier $e_i$ is internal only: it builds structural descriptors and is never emitted.

For internal code paths (e.g., direct branches and internal calls), the module resolves $x_i=\texttt{addr}$ and records the destination target in $e_i$. Opcode-side anchors preserve the opcode text while binding the current execution location via $e_i$. External API stubs retain their stable imported symbols (e.g., \texttt{printf@plt}) without consuming an internal descriptor. Data references are canonicalized by section permissions, selecting $x_i$ from
\[
\{\texttt{daddr},\texttt{daddr\_rodata},\texttt{daddr\_bss},\texttt{daddr\_unknown}\},
\]
while assigning the absolute memory reference to $e_i$. Stack-relative variables map to \texttt{var}, with their frame-relative displacement stored in $o_i$. Finally, standard constants and unresolved pointer candidates default conservatively to $\texttt{imm}$ and $\texttt{num}$, respectively, while non-numeric tokens set $e_i=o_i=\bot$.

Crucially, co-referent code and data references consistently reuse the same internal entity identifier $e_i$, while stack-relative references reuse the same signed displacement $o_i$ across instructions. Raw numeric literals are therefore excluded from lexical vocabulary expansion and are restricted solely to background descriptor generation.

\subsection{Structural Descriptor Construction}
\label{sec:design_encoding}

Following typing and canonicalization, \mytool\ formalizes grounded entities into type-specific, normalized hierarchical descriptors, while stack entities bypass layout normalization to retain raw signed displacements. Formally, the coordinate descriptor $p_i$ is mapped via:
$$
p_i =
\begin{cases}
 p_i^{\text{code}} \in [0,1]^3, & r_i=\mathrm{code} \\
 p_i^{\text{data}} \in [0,1]^2, & r_i=\mathrm{data} \\
 \bot, & \text{otherwise}.
\end{cases}
$$
Descriptors are generated only when the enclosing spans are valid and non-degenerate; otherwise, the structural route is masked to avoid introducing unsupported topology.

\vspace{2pt}
\noindent
\textbf{\textit{Code structure and Opcode-Side Anchoring}.}
Let a code entity resolve to a virtual address $a$. For opcode-side anchors, $a$ marks the current execution location; for control-flow targets, $a$ designates the resolved destination. Given that $a$ resides within basic block $[b_s,b_e)$, function $[f_s,f_e)$, and total binary span $[B_{\min},B_{\max})$, \mytool\ captures its cross-boundary position via a three-tier hierarchical descriptor $p_i^{\text{code}}= \big( p_i^{\mathrm{func}}, p_i^{\mathrm{bb}}, p_i^{\mathrm{inst}} \big)$, where:
$$
p_i^{\mathrm{func}}
=
\frac{f_s - B_{\min}}{B_{\max} - B_{\min}},~~ 
p_i^{\mathrm{bb}}
=
\frac{b_s - f_s}{f_e - f_s},~~
p_i^{\mathrm{inst}}
=
\frac{a - b_s}{b_e - b_s}.
$$
\textbf{\textit{Shared Coordinate System}.}
This multi-tier formulation exposes structural relations that are difficult to recover from token order alone. Unlike prior work that treats jump targets as isolated symbolic tokens, \mytool\ places both the source instruction and its destination target in a shared spatial coordinate system. For example, consider the \texttt{jle 0x1185} instruction at \texttt{0x1176} in Figure~\ref{fig:systemdesign} (Step B). \mytool\ anchors the opcode token \texttt{jle} to the source address \texttt{0x1176}, yielding the source-side coordinate $(0.27, 0.00, 0.92)$, while canonicalizing the target operand as \texttt{addr} and grounding it to \texttt{0x1185}, yielding the destination-side coordinate $(0.27, 0.44, 0.00)$. Exposing these paired coordinates at the input boundary allows the ALM to model relative spatial relations between a branch instruction and its target, even when they are distant in the token sequence. External library calls use no internal code descriptor and set $p_i=\bot$.

\vspace{2pt}
\noindent
\textbf{\textit{Data structure}.}
Data entities encode section-level locality within the binary layout. Assuming a static data pointer targets address $a_d$ within a specific section $[s_s,s_e)$, its spatial orientation is bounded by a two-tier descriptor $p_i^{\text{data}}=\big( p_i^{\mathrm{sec}}, p_i^{\mathrm{obj}} \big)$, calculated as:
$$
p_i^{\mathrm{sec}}
=
\frac{s_s - B_{\min}}{B_{\max} - B_{\min}},
\qquad
p_i^{\mathrm{obj}}
=
\frac{a_d - s_s}{s_e - s_s}.
$$
While the emitted canonical token (e.g., \texttt{daddr\_rodata}) preserves coarse section permissions, $p_i^{\text{data}}$ provides finer-grained normalized locality beyond section-level typing. For example, as shown in Figure~\ref{fig:systemdesign}, 
a resolved data operand (\texttt{daddr}) receives a normalized descriptor such as $(0.49, 0.14)$, providing a compact locality cue beyond the section-level token.
If section attributes are unavailable or unreliable, \mytool{} sets $p_i=\bot$.

\vspace{2pt}
\noindent
\textbf{\textit{Stack structure}.}
Because local stack slots lack static absolute anchors within the binary image, mapping them to global binary coordinates would distort their layout semantics. \mytool\ therefore preserves their structural identity purely via the frame-relative signed offset $o_i=o$ extracted during canonicalization, keeping $p_i=\bot$. As illustrated in Figure~\ref{fig:systemdesign}, a stack operand (\texttt{var}) does not use 3D layout normalization and instead retains its frame-relative geometry (e.g., $-20$).

\subsection{Host-Compatible Serialization}
\label{sec:design_serialization}
Following descriptor construction, NumAnchor serializes each sample into the precise input carrier prescribed by the host ALM—whether instantiated as localized instruction pairs, basic-block control flow paths, or function-level token streams augmented with native execution traces. Irrespective of the carrier topology, each serialized element retains the address-aware structural tuple:\[(x_i,r_i,p_i,o_i,z_i),\]where $x_i$ is the model-facing lexical token, $r_i$ is the semantic routing path, $p_i$ is the structural coordinate descriptor, $o_i$ is the signed stack-relative offset, and $z_i$ encapsulates host-specific auxiliary fields. As conceptualized in Figure~\ref{fig:systemdesign}, this architecture establishes a multi-channel alignment protocol: the lexical row follows the host model's native vocabulary stream, while parallel side channels carry route-specific structural geometries, automatically masking inactive routing paths with $\bot$.

This decoupled design keeps the host-facing sequence compact. By offloading raw numeric values from the primary token stream into parallel coordinate descriptors, NumAnchor avoids adding lexical tokens and mitigates vocabulary sparsity caused by raw numeric literals. Consequently, the host Transformer backbone receives its native instruction tokens alongside structurally-grounded side fields simultaneously at the representation boundary. For execution-aware or dynamic-tracing backbones, this structural augmentation interfaces exclusively with the static token-side channels, without modifying dynamic-trace or state channels.

\subsection{Input-Side Fusion and Structural Lifting}
\label{sec:design_fusion}

Let $h_i^{\mathrm{host}}$ be the baseline input representation synthesized by the original host model frontend (e.g., a BERT-style token initialization):
\[
h_i^{\mathrm{host}} = E_{\mathrm{tok}}(x_i) + E_{\mathrm{pos}}(i) + E_{\mathrm{seg}}(s_i) + E_{\mathrm{side}}(z_i),
\]
where $E_{\mathrm{side}}$ is omitted natively if auxiliary channels are absent. NumAnchor injects structure-aware grounding directly at the token boundary prior to the first self-attention layer:
\[
\begin{aligned}
h_i^{(0)}
= h_i^{\mathrm{host}}
&+ \operatorname{Ind}\!\left[r_i=\mathrm{code}\land p_i\neq\bot\right]\, E_{\mathrm{code}}\!\left(p_i^{\mathrm{code}}\right)\\
&+ \operatorname{Ind}\!\left[r_i=\mathrm{data}\land p_i\neq\bot\right]\, E_{\mathrm{data}}\!\left(p_i^{\mathrm{data}}\right)\\
&+ \operatorname{Ind}\!\left[r_i=\mathrm{stack}\land o_i\neq\bot\right]\, E_{\mathrm{stack}}\!\left(o_i\right),
\end{aligned}
\]
where $\operatorname{Ind}[\cdot]\in\{0,1\}$ is the indicator function. 
Elements routed to $\mathrm{imm}$ or $\mathrm{num}$, or those tagged with $\bot$ side fields, rely entirely on $h_i^{\mathrm{host}}$, while grounded entities receive route-specific structural signals.

\vspace{2pt}
\noindent
\textbf{\textit{High-Dimensional Fourier Lifting}.}
At the code and data routes, \mytool{} feeds normalized layout coordinates to lightweight multilayer perceptron (MLP) projections. Using each coordinate as a single scalar can bias these projections toward smooth, low-frequency patterns, limiting their ability to capture fine-grained positional variation~\cite{fourier19,MLPfourier20}. To provide a multi-scale representation, \mytool{} lifts each 1D continuous coordinate $u\in[0,1]$ into a periodic feature space using a Fourier mapping $\phi(u;s) \in \mathbb{R}^{d_F}$, where $d_F$ is an even feature dimension and $K=d_F/2$:
\begin{equation}
\phi(u;s) = \big[ \sin(u \cdot s \cdot \omega_k), \cos(u \cdot s \cdot \omega_k) \big]_{k=0}^{K-1},
\end{equation}
with $\omega_k = 10000^{-2k/d_F}$, and $s>0$ serving as a frequency-scale hyperparameter instantiated separately for code ($s_c$) and data ($s_d$) domains. For brevity, we denote $\phi_s(\cdot) = \phi(\cdot; s)$. 

\vspace{2pt}
\noindent
\textbf{\textit{Effect of Fourier Lifting}.} 
Fourier lifting provides a multi-scale basis for normalized layout coordinates~\cite{fourier19,MLPfourier20}. Code and data locality can matter at different granularities: coarse coordinates indicate broad function, block, or section regions, while fine-grained offsets distinguish nearby instruction or object locations. By mapping $u$ across multi-frequency sinusoids, low-frequency components capture coarse layout information, while higher-frequency components preserve sensitivity to fine-grained locality, such as instruction-level offsets. This gives the route-specific projection access to both coarse and fine address-grounded structure, rather than forcing it to rely on a single scalar coordinate.

The code and data structural projections are subsequently formulated via separate MLPs:
\[
E_{\mathrm{code}}\!\left(p_i^{\mathrm{code}}\right) = \operatorname{MLP}_{\mathrm{code}}\!\left( \phi_{s_c}\!\!\left(p_i^{\mathrm{func}}\right) \mathbin{\Vert} \phi_{s_c}\!\!\left(p_i^{\mathrm{bb}}\right) \mathbin{\Vert} \phi_{s_c}\!\!\left(p_i^{\mathrm{inst}}\right) \right),
\]
\[
E_{\mathrm{data}}\!\left(p_i^{\mathrm{data}}\right) = \operatorname{MLP}_{\mathrm{data}}\!\left( \phi_{s_d}\!\!\left(p_i^{\mathrm{sec}}\right) \mathbin{\Vert} \phi_{s_d}\!\!\left(p_i^{\mathrm{obj}}\right) \right),
\]
where $\Vert$ indicates feature concatenation, satisfying $\operatorname{MLP}_{\mathrm{code}}:\mathbb{R}^{3d_F}\to\mathbb{R}^{d_h}$ and $\operatorname{MLP}_{\mathrm{data}}:\mathbb{R}^{2d_F}\to\mathbb{R}^{d_h}$. Decoupling these mappings is essential, as code hierarchy and static data locality capture fundamentally distinct execution behaviors.

\vspace{2pt}
\noindent
\textbf{\textit{Stack Offset Encoding}.}
Unlike code or static data, stack slots lack absolute layouts within the binary image. NumAnchor uses a hybrid discrete-continuous encoding for stack variables. Setting $\tau=512$ (covering approximately 92\% of empirical \texttt{var} occurrences within the StateFormer training distribution), valid displacements $o_i\neq\bot$ are categorized into precise or overflow buckets via $\beta(o_i)$:
\[
\beta(o_i)=
\begin{cases}
 o_i+\tau, & -\tau \le o_i \le \tau,\\
 2\tau+1, & o_i > \tau,\\
 2\tau+2, & o_i < -\tau.
\end{cases}
\]
The stack representation $E_{\mathrm{stack}}(o_i)$ merges localized categorical identity with global continuous scales:
\[
E_{\mathrm{stack}}\!\left(o_i\right) = \operatorname{Emb}_{\mathrm{off}}\!\left(\beta(o_i)\right) + \operatorname{MLP}_{\mathrm{dist}}\!\left( \log(1+|o_i|) \right).
\]
Here, $\operatorname{Emb}_{\mathrm{off}}:\{0,\ldots,2\tau+2\}\to\mathbb{R}^{d_h}$ maps bucketed indicators to the hidden space, while $\operatorname{MLP}_{\mathrm{dist}}:\mathbb{R}\to\mathbb{R}^{d_h}$ projects the logarithmic distance feature. This hybrid design supports exact bucket-level matching for frequent local stack slots, while preserving coarse magnitude markers for rare, extreme frame displacements without inducing excessive embedding-table growth.

\subsection{Portability and Fallback Design}
\label{sec:design_generalization_tradeoffs}


NumAnchor operates at the input boundary, adding structural embeddings before the first Transformer layer without changing the host backbone, prediction heads, or training objectives. This design supports different ALM input formats. In our experiments, NumAnchor is trained jointly with each host model during its regular pre-training, without introducing an additional pre-training objective or a separate pre-training stage. When static analysis cannot reliably ground a numeric operand as a code reference, data reference, or stack-relative value, NumAnchor masks the corresponding structural side fields and falls back to generic \texttt{imm} or \texttt{num} tokens.

\FloatBarrier
\FloatBarrier

\section{Evaluation}
\label{sec:evaluation}

\begin{table}[H]
\centering
\caption{Operand-level statistics in stripped binaries. Percentages are measured over all operand occurrences; Code, Data, Stack, Imm., and Other sum to Num./All.}

\label{tab:numeric_operand_stats}
\resizebox{0.49\textwidth}{!}{
\begin{tabular}{lrrrrrr}
\toprule
Dataset & Num./All & Code & Data & Stack & Imm. & Other \\
\midrule
StateFormer & 45.73\% & 14.41\% & 4.87\% & 12.47\% & 12.11\% & 1.87\% \\
jTrans      & 40.93\% & 9.97\%  & 8.22\% & 9.17\%  & 7.63\%  & 5.94\% \\
\bottomrule
\end{tabular}
}

\end{table}

\begin{table} [H]
\centering
\small
\caption{Evaluated models. Pre-training tasks: MLM, CWP, DUP, JTP and GSM refer to Masked Language Modeling, Context Window Prediction, Def-Use Prediction, Jump Target Prediction, Generative State Modeling, respectively.}

\label{tab:eval_models}
\resizebox{0.49\textwidth}{!}{
\begin{tabular}{@{}p{1.85cm}p{1.65cm}p{2.05cm}p{3.0cm}@{}}
\toprule
\textbf{Model} & \textbf{Family} & \textbf{Pre-training tasks} & \textbf{Evaluation tasks} \\
\midrule
 
PalmTree~\cite{li2021palmtree}
& Instruction-level
& MLM, CWP, DUP
& Intrinsic probes; func. similarity search \\
 
BERT~\cite{raid25}
& Function-level
& MLM
& Func. similarity search \\
 
jTrans~\cite{wang2022jtrans}
& Function-level
& MLM, JTP
& Func. similarity search \\
 
\textsc{jTrans-CLAP}~\cite{wang2024clap}
& Function-level
& MLM, JTP
& Func. similarity search \\
 
StateFormer~\cite{pei2021stateformer}
& Function/code-block-level
& GSM
& Type inference \\
 
TREX~\cite{pei2020trex}
& Function-level
& MLM
& Func. similarity search \\
\bottomrule
\end{tabular}
}

\end{table}

\subsection{Experimental Setup}
\label{sec:eval_setup}
We use a controlled within-model evaluation protocol to isolate the impact of \mytool. Each baseline is compared only with its \mytool-enhanced variant, while keeping the neural architecture, training hyperparameters, and task formulation unchanged. The main paper reports all primary results needed to answer the RQs.

\vspace{2pt}
\noindent
\textbf{Benchmarks and Empirical Motivation.}
Models are evaluated on their prescribed benchmarks: \textit{BinaryCorp-3M} for the jTrans lineage and the \textit{StateFormer dataset} for the remaining baselines. 
\textit{BinaryCorp-3M} contains $10,265$ binaries and about 3.6M functions, while the \textit{StateFormer dataset} includes 33 x86-64 open-source projects under multiple optimization settings. Numeric operands account for 45.73\% of operand occurrences in StateFormer and 40.93\% in jTrans (Table~\ref{tab:numeric_operand_stats}), highlighting numeric ambiguity as a key challenge for binary representation learning.

\vspace{2pt}
\noindent
\textbf{Preprocessing and Controlled Integration.} Inputs are reconstructed using IDA Pro~\cite{hexrays_idapro}, with UniCorn~\cite{quynh2015unicorn} regenerating dynamic micro-traces strictly for execution-aware models. Our goal is to isolate how different numeric/address encoding strategies affect ALM representations under controlled host settings; the evaluated host models, input granularities, pre-training tasks, and downstream tasks are summarized in Table~\ref{tab:eval_models}. Accordingly, \mytool\ is integrated at the representation boundary while preserving each host model's native input granularity: token-level rewriting for PalmTree and BERT, function-level WordPiece replacement for jTrans and \textsc{jTrans-CLAP}, and static token-side substitution for StateFormer and TREX. For \textsc{jTrans-CLAP}, we do not reproduce the full CLAP code--text contrastive pipeline because its LLM-assisted supervision corpus and generation/filtering pipeline are not publicly released in a reusable form. 
Instead, we implement CLAP's reproducible assembly-side components---address rebasing, WordPiece tokenization, and instruction embeddings for boundary preservation---on the original jTrans backbone and training pipeline.
This variant serves as a controlled comparison to CLAP-style address processing.

\vspace{2pt}
\noindent
\textbf{Evaluation Metrics.}
Performance is measured via Recall@$k$ and Mean Reciprocal Rank (MRR) for function similarity retrieval, and non-\texttt{NA} Precision, Recall, and F1 scores for type recovery. 

\vspace{4pt}
Taken together, this tightly controlled experimental framework allows us to systematically evaluate \mytool\ across four complementary Research Questions (RQs):
\begin{itemize}
    \item \textbf{RQ1 (Structural Recoverability):} Does \mytool make static-analysis-recovered program structure recoverable from learned ALM representations?
    \item \textbf{RQ2 (Downstream Effectiveness):} Do \mytool-induced representation gains transfer across downstream binary-analysis tasks and ALM families?
    \item \textbf{RQ3 (Structural Channel Ablation):} Which structural channels, including code, data, and stack descriptors, contribute most to the observed gains?
    \item \textbf{RQ4 (Engineering Overhead):}
    What are the parameter, training, memory, and inference
    overheads of \mytool{}?
\end{itemize}

\subsection{Probing Address Structure (RQ1)}
\label{sec:rq1}

We use PalmTree as an instruction-centric testbed to examine whether the address structure introduced by \mytool\ remains visible in frozen instruction embeddings. We compare the original PalmTree baseline, a NAU-based address-aware variant, and full \mytool. 

We first evaluate a hierarchical-location probe, which predicts an instruction's relative bucket within the binary, function, and basic block. The original PalmTree embeddings provide only limited hierarchical locality information, achieving 28.3\%, 30.1\%, and 42.6\% accuracy. The NAU variant substantially improves binary-level position prediction to 83.0\%, but provides little benefit for function- and block-level locality, reaching only 31.2\% and 44.1\%. In contrast, full \mytool\ consistently achieves high accuracy across all three granularities, reaching 91.3\%, 95.7\%, and 98.7\%, respectively.


We further evaluate whether control-flow destinations are reflected in the embedding space. The baseline is close to random for both jump and call targets, with AUROC scores of 0.496 and 0.504. The NAU variant improves alignment to 0.821 and 0.784, while full \mytool\ further increases performance to 0.932 and 0.907. Together with the hierarchical-location results, these findings show that the structural information introduced by \mytool{} remains recoverable from frozen instruction embeddings after pre-training and contextualization. Since the probed properties are directly encoded in the input descriptors, RQ1 measures information retention rather than the learning of additional program relations. Our downstream ablations provide complementary evidence for the utility of code, data, and stack descriptors, while probing indirect relations such as def-use remains future work.

\subsection{Downstream Evidence Across ALM Families (RQ2)}
\label{sec:rq23}

We next evaluate whether the representation-level improvements observed in Section~\ref{sec:rq1} are reflected in downstream binary-analysis tasks, and whether \mytool\  provides consistent within-model gains across different ALM families. Since the evaluated models differ in input granularity, preprocessing strategy, auxiliary signals, and downstream objectives, all results in this subsection should be interpreted as baseline-versus-\mytool\  comparisons within each host model, rather than as direct cross-model rankings.

\begin{figure}[t]
    \centering
    \includegraphics[width=0.98\columnwidth]{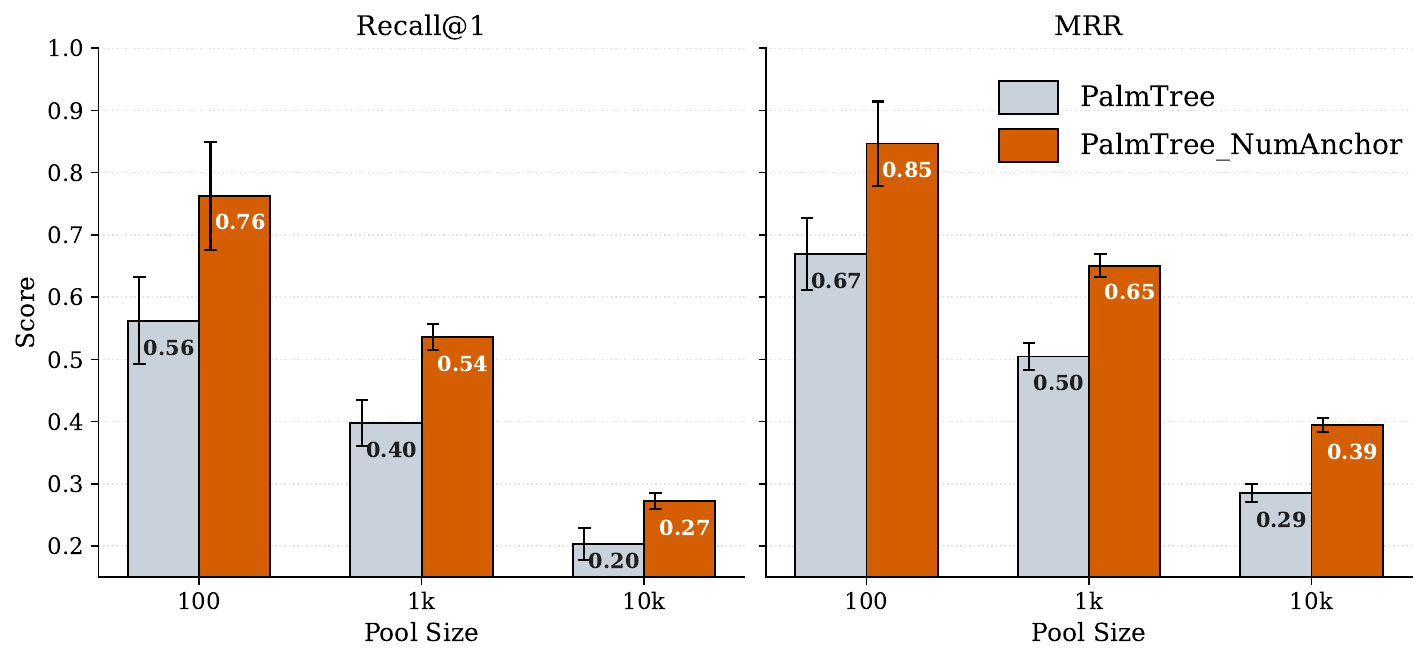}
    \vspace{-2mm}
    \caption{Function-similarity retrieval performance of PalmTree and PalmTree\_\mytool\  on the StateFormer-x64 dataset under different retrieval pool sizes. Bars represent the mean scores across repeated sampling trials, and error bars indicate the corresponding standard deviation ($\pm\mathrm{std}$).}
    \label{fig:funcsimPalmtree}

\end{figure}

\begin{figure}[t]
    \centering
    \includegraphics[width=0.98\columnwidth]{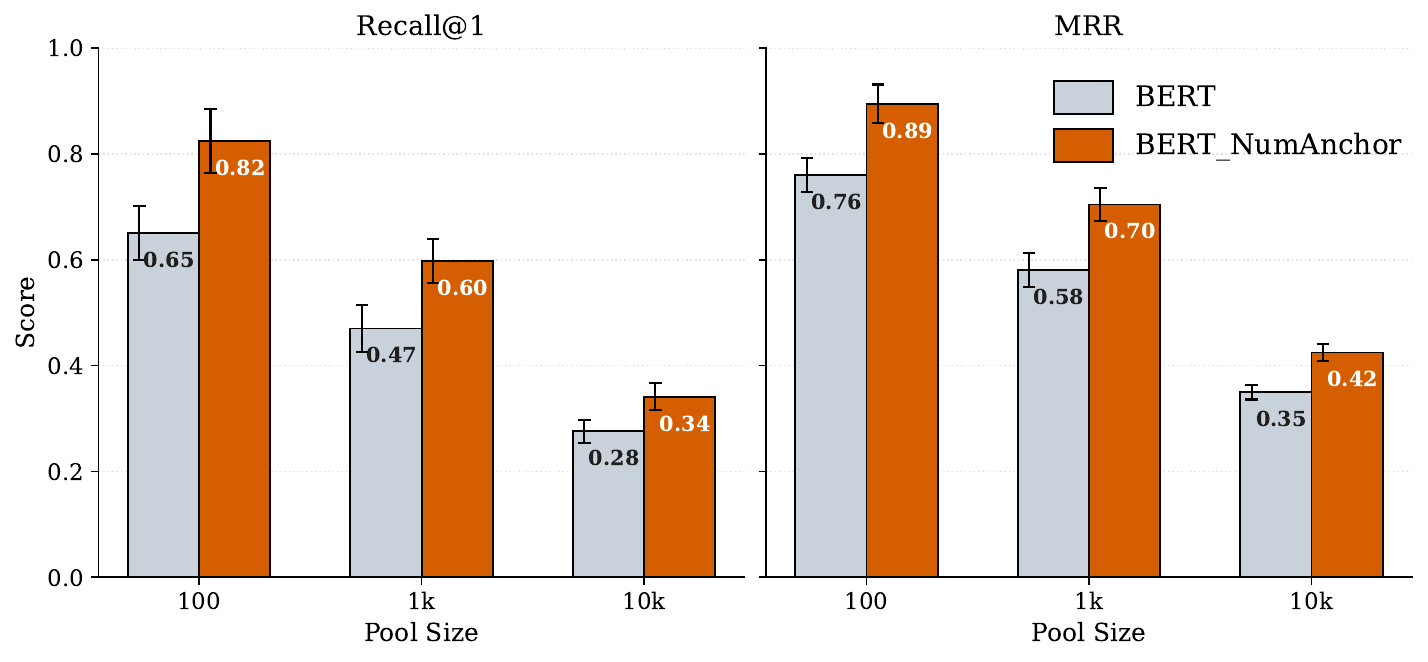}
    \vspace{-2mm}
    \caption{Function-similarity retrieval performance of BERT and BERT\_\mytool\  on the StateFormer-x64 dataset under different retrieval pool sizes.}
    \label{fig:funcsimBert}
\end{figure}

\vspace{2pt}
\noindent
\textbf{\textit{Static-Code Carriers: PalmTree~\cite{li2021palmtree} and Vanilla BERT~\cite{raid25}.}}
We evaluate \mytool{} on two static-code carriers with different input granularities: instruction-level aggregation in PalmTree and whole-function sequence tokenization in Vanilla BERT. 
Following established evaluation protocols~\cite{raid25}, each query is retrieved from candidate pools of 100, 1k, and 10k functions, and we report mean metrics over 20 randomized trials (Figures~\ref{fig:funcsimPalmtree} and~\ref{fig:funcsimBert}).

The results show that \mytool{} consistently improves both carriers across candidate-pool sizes. As the pool grows, all methods face a harder retrieval problem because more distractor functions are introduced. However, the \mytool{}-enhanced variants degrade more slowly than their corresponding baselines, indicating that the added structural descriptors help preserve discriminative information at larger retrieval scales. In the 10k-candidate setting, \mytool{} improves PalmTree by 35.0\% in Recall@1 and 31.7\% in MRR, and improves Vanilla BERT by approximately 21\% and 20\%, respectively. These results suggest that representing numeric-like operands as typed structural descriptors benefits both fine-grained instruction carriers and coarse sequential encoders, helping the resulting representations retain structurally relevant signals and distinguish true cross-optimization matches from increasingly large sets of distractors.

\begin{table*}[t]
\centering
\caption{Cross-optimization function retrieval performance of jTrans variants on 10k-candidate pools constructed from the jTrans BinaryCorp-3M dataset. Each query ranks one ground-truth match against 9,999 deduplicated distractor functions.}

\label{tab:jtrans-local-10k}
\setlength{\tabcolsep}{4pt}
\renewcommand{\arraystretch}{1.12}
\resizebox{0.95\textwidth}{!}{%
\begin{tabular}{@{}lccccccc@{\hspace{10pt}}ccccccc@{}}
\toprule
& \multicolumn{7}{c}{MRR $\uparrow$} & \multicolumn{7}{c}{Recall@1 $\uparrow$} \\
\cmidrule(lr){2-8}\cmidrule(lr){9-15}
Model
& O0/O3 & O1/O3 & O2/O3 & O0/Os & O1/Os & O2/Os & Avg.
& O0/O3 & O1/O3 & O2/O3 & O0/Os & O1/Os & O2/Os & Avg. \\
\midrule
jTrans
& 0.422 & 0.690 & 0.744 & 0.482 & 0.671 & 0.710 & 0.620
& 0.321 & 0.611 & 0.653 & 0.412 & 0.594 & 0.629 & 0.537 \\
jTrans\_\mytool\ 
& \textbf{0.498} & \textbf{0.749} & \textbf{0.860} & \textbf{0.587} & \textbf{0.774} & \textbf{0.767} & \textbf{0.706}
& \textbf{0.395} & \textbf{0.667} & \textbf{0.803} & \textbf{0.477} & \textbf{0.688} & \textbf{0.684} & \textbf{0.619} \\

\midrule[1.2pt]

JTrans-CLAP
& 0.377 & 0.638 & 0.696 & 0.425 & 0.608 & 0.667 & 0.569
& 0.298 & 0.569 & 0.640 & 0.350 & 0.516 & 0.596 & 0.495 \\

JTrans-CLAP\_\mytool\ 
& \textbf{0.501} & \textbf{0.753} & \textbf{0.872} & \textbf{0.581} & \textbf{0.732} & \textbf{0.721} & \textbf{0.693}
& \textbf{0.392} & \textbf{0.671} & \textbf{0.813} & \textbf{0.468} & \textbf{0.653} & \textbf{0.645} & \textbf{0.607} \\
\bottomrule
\end{tabular}%
}

\end{table*}

\vspace{2pt}
\noindent
\textbf{\textit{Address-Rebasing Family: jTrans~\cite{wang2022jtrans} and \textsc{jTrans-CLAP~\cite{wang2024clap}}.}}
We next evaluate \mytool{} in a rebasing-oriented, function-level retrieval setting. Both jTrans and the adapted \textsc{jTrans-CLAP} incorporate address-sensitive mechanisms, but their explicit alignment mainly targets localized control-flow jumps. \mytool{} therefore provides a controlled way to replace or augment these address interfaces with a unified structural representation across code, data, and stack operands.
As shown in Table~\ref{tab:jtrans-local-10k}, \mytool{} improves both baselines across all six cross-optimization retrieval pairs. It increases jTrans average MRR/Recall@1 from 0.620/0.537 to 0.706/0.619, and yields larger gains for \textsc{jTrans-CLAP}, whose average MRR/Recall@1 improves from 0.569/0.495 to 0.693/0.607. These results suggest two observations.


\textbf{First}, the gains indicate that address modeling benefits from structural information beyond localized jump-target rebasing. \textbf{Second}, the lower performance of vanilla \textsc{jTrans-CLAP} compared with jTrans suggests that finer-grained WordPiece tokenization and sequence-relative rebasing alone do not necessarily yield stable representations under cross-optimization changes. After adding \mytool{}'s structural descriptors, \textsc{jTrans-CLAP} recovers much of this gap and approaches the performance of \texttt{jTrans\_\mytool}. This supports our claim that typed, multi-level structural locality provides a more reliable address representation interface than relying solely on lexical sub-token overlap or local rebasing heuristics.

\vspace{2pt}
\noindent
\textbf{\textit{Execution-Aware Family: TREX~\cite{pei2020trex} and StateFormer~\cite{pei2021stateformer}.}}
Execution-aware architectures provide a complementarity test because they already incorporate dynamic traces, register states, or memory states. We therefore do not expect a static representation to replace these dynamic signals. Instead, we evaluate whether grounded numeric semantics provide additional structural information while leaving the host execution-aware components unchanged.


The results show that \mytool{} provides complementary gains. In the 10k-candidate retrieval setting, TREX Recall@1/MRR improves from 0.28/0.40 to 0.31/0.43, corresponding to a 10.7\% relative Recall@1 gain. On StateFormer's native type-recovery task, \mytool{} improves F1 at every optimization level, with gains of +5.2, +3.6, +2.9, and +2.0 points from O0 to O3.

\vspace{2pt}
\noindent
\textbf{\textit{RQ2 takeaway.}}
Overall, these downstream results show that \mytool{} improves ALM performance across both instruction-level and function-level input granularities, spanning static-code, address-rebasing, and execution-aware settings under matched within-backbone comparisons. 
We next analyze the contributing components.

\subsection{Ablation Study of \mytool\ (RQ3)}
\label{sec:ablation}

\begin{table}[t]
\centering
\caption{Ablation configurations (top) and evaluation results (bottom) for PalmTree on 10k-candidate retrieval. \textsc{Role-only}: type tags without spatial coordinates; \textsc{NAU}: baseline continuous numeric encoder; \textsc{Fourier} \& \textsc{Struct. Enc.}: coordinate lifting and type-specific projections. $\Delta$ indicates absolute improvement over the vanilla baseline; best results are bolded.}
\label{tab:palmtree_ablation_combined}
\renewcommand{\arraystretch}{1.15}

\resizebox{\columnwidth}{!}{
\begin{tabular}{lcccccc}
\toprule
\textbf{Variant} & \textbf{Code} & \textbf{Data} & \textbf{Stack} & \textbf{NAU} & \textbf{Fourier} & \textbf{Struct. Enc.} \\
\midrule
PalmTree              &              &              &              &              &              &              \\
Code-only             & $\checkmark$ &              &              &              & $\checkmark$ & $\checkmark$ \\
Code+Data             & $\checkmark$ & $\checkmark$ &              &              & $\checkmark$ & $\checkmark$ \\
Role-only             & $\checkmark$ & $\checkmark$ & $\checkmark$ &              &              &              \\
NAU variant           & $\checkmark$ & $\checkmark$ & $\checkmark$ & $\checkmark$ &              &              \\
NumAnchor w/o Fourier & $\checkmark$ & $\checkmark$ & $\checkmark$ &              &              & $\checkmark$ \\
NumAnchor             & $\checkmark$ & $\checkmark$ & $\checkmark$ &              & $\checkmark$ & $\checkmark$ \\
\bottomrule
\end{tabular}
}

\vspace{6pt} 

\resizebox{\columnwidth}{!}{
\begin{tabular}{lcccc}
\toprule
\textbf{Variant} & \textbf{MRR} & \textbf{$\Delta$MRR} & \textbf{Recall@1} & \textbf{$\Delta$R@1} \\
\midrule
PalmTree              & 0.293 & -- & 0.203 & -- \\
Code-only             & 0.321 & +0.028 & 0.245 & +0.042 \\
Code+Data             & 0.338 & +0.045 & 0.256 & +0.053 \\
Role-only             & 0.305 & +0.012 & 0.229 & +0.026 \\
NAU variant           & 0.339 & +0.046 & 0.241 & +0.038 \\
NumAnchor w/o Fourier & 0.346 & +0.053 & 0.248 & +0.045 \\
NumAnchor             & \textbf{0.386} & \textbf{+0.093} & \textbf{0.274} & \textbf{+0.071} \\
\bottomrule
\end{tabular}
}

\end{table}

\vspace{2pt}
\noindent
\textbf{\textit{Ablation Study and Component Analysis.}}
To isolate the contribution of each design component, we conduct component ablations under the $10\mathrm{k}$-candidate PalmTree retrieval setting, as summarized in Table~\ref{tab:palmtree_ablation_combined}. The full \mytool{} design achieves the best performance, improving Recall@1 and MRR by 35.0\% and 31.7\% relative to PalmTree, respectively. In absolute terms, the +0.071 Recall@1 gain means that the correct function is ranked first for about seven additional queries per hundred in this large-candidate retrieval setting.

The PalmTree ablation results provide two main observations. \textbf{First, code, data, and stack descriptors contribute complementary gains.} Adding \textsc{Code} coordinates already improves Recall@1 by +0.042, showing that explicit code-location information helps the model recover control-flow-related structure. Adding \textsc{Data} and then \textsc{Stack} descriptors further increases the gain to +0.071, indicating that memory references and stack-frame layouts provide useful semantic anchors beyond control-flow targets alone. This supports our hypothesis that numeric-like operands carry structural information beyond code addresses.


\textbf{Second, explicit structural encoding is more effective than role tags or generic numeric modeling.} The \textsc{Role-only} variant exposes code/data/stack types but does not encode their structural locations, while the \textsc{NAU variant} treats numeric values through a continuous numeric encoder. Both variants trail the full design, suggesting that operand type alone or scalar numeric regression is insufficient to capture the locality and entity relationships needed for retrieval. The \textsc{NumAnchor w/o Fourier} variant improves over the NAU variant, but full \mytool{} further raises MRR from 0.346 to 0.386, indicating that Fourier lifting contributes additional signal beyond the structural routes alone. Overall, the structural encoder---with Fourier lifting for code/data coordinates and type-specific stack encoding---provides a better inductive bias than generic numeric modeling.

We further observe the same additive trend on the jTrans backbone. Code descriptors improve the jTrans baseline, data descriptors provide further gains, and the full \mytool{} variant performs best, improving Avg. MRR/Recall@1 from 0.620/0.537 to 0.706/0.619. Since jTrans already uses jump-aware symbolic rebasing, these results suggest that typed structural grounding complements existing address heuristics rather than merely duplicating them.



\subsection{Overhead of \mytool\ (RQ4)}
\label{sec:overhead}
\noindent
\mytool\ adds side-channel encoders but does not increase the host sequence length.
For a 768-dimensional PalmTree backbone, it adds 4.73M parameters over an 86.31M-parameter baseline, a 5.48\% increase.
For PalmTree, the training cost increases from 1.29 to 1.32 GPU-hours per epoch (2.3\%), while peak memory usage increases from 38.32 to 42.65 GB (11.3\%).
Inference overhead is also small: over groups of 20 functions, \mytool{} adds 0.15s for PalmTree, 0.11s for jTrans, and 0.08s for TREX, corresponding to at most about 7.5ms per function.

\section{Threats to Validity}
\label{sec:discussion_threats}

This section clarifies the scope of our evidence and summarizes the main implications. These threats affect how the results should be interpreted, but do not change the controlled within-model comparisons used throughout the evaluation.

\vspace{2pt}
\noindent
\textbf{\textit{Construct Validity and Alternative Paradigms.}}
We use function retrieval and type recovery as proxy tasks for evaluating binary-code representations. Although not exhaustive, they capture core capabilities for semantic matching, layout-aware representation learning, and type-related reasoning. \mytool{} complements decompiled-pseudocode approaches by preserving address-grounded assembly-level signals, including code locality, stack-frame offsets, and data-section structure, that may be abstracted away during lifting. However, \mytool{} is limited to statically observable information and maps unresolved heap offsets or complex pointer-derived addresses to generic numeric tokens. We reduce evaluation bias by following standard retrieval/type-recovery metrics and the TYGR deduplication protocol~\cite{zhu2024tygr}.



\vspace{2pt}
\noindent
\textbf{Internal Validity and Boundary Dependence.}
We evaluate all variants under fixed architectures, training settings, and task formulations to isolate the contribution of \mytool{}. Our prototype relies on IDA Pro~\cite{hexrays_idapro} to recover function boundaries and related artifacts, consistent with the preprocessing required by the evaluated ALMs. Incorrect function boundaries can distort function-relative descriptors and host input units defined over recovered functions, while local instruction and section-level signals are less directly affected. Incorrect cross-references may also associate unrelated operands with the same entity, introducing misleading structural information. When grounding information is unavailable or identified as unreliable, \mytool{} masks the corresponding descriptors. However, it cannot reliably detect or correct erroneous structural information that the analysis frontend accepts as valid, which may affect downstream performance.

\vspace{2pt}
\noindent
\textbf{Layout Correlation in Function Retrieval.}
Our project-level data splits and function deduplication reduce data overlap but do not explicitly control for layout correlations across optimization levels. Since \mytool{} encodes hierarchical code locations, similarities in function or basic-block layouts may contribute to the observed retrieval gains. Improvements in the separate type-recovery task provide complementary evidence for the utility of structural information, but do not rule out possible layout-related effects on retrieval performance. Further evaluation, such as removing binary-level function coordinates or randomizing layouts, is left for future work.



\vspace{2pt}
\noindent
\textbf{External Validity and Obfuscation.}
Our evaluation focuses on x86/x64 binaries drawn from standard benchmark corpora under standard compiler optimizations, where \mytool{} shows consistent gains across multiple ALM families and downstream settings. We have not yet evaluated other ISAs such as ARM or MIPS, which require comparable static-analysis artifacts and ISA-specific validation. The underlying approach may extend to other code representations, but adapting it beyond assembly would require structural descriptors suited to those representations. Strong obfuscation may also disrupt the static topography used by \mytool{}, especially when control-flow layout or data-reference structure is intentionally obscured. We leave cross-ISA evaluation, extensions beyond assembly, integration with learning-based indirect control-flow recovery methods~\cite{ICFlowNet,NeuCall}, and robustness under stronger obfuscation as future work.




\section{Conclusion}
\label{sec:conclusion}

We presented NumAnchor, a structure-aware input-boundary
interface that represents numeric-like operands using typed,
address-grounded structural information. By preserving
recoverable code, data, and stack information at the model
input boundary, NumAnchor enables ALMs to better capture
the structural semantics and relationships encoded in numeric
operands. This design improves representation quality and
downstream performance across multiple ALM families without
modifying their backbone architectures or training objectives.
Our experimental results demonstrate the effectiveness and
generalizability of this approach across diverse binary analysis
tasks, highlighting the practical value of exposing recovered
program structure to ALMs through numeric operand
representations. These findings suggest that incorporating
address-grounded structural information at the input boundary
offers a practical and broadly applicable direction for
improving assembly language representations.


\section*{Data Availability}

To facilitate reproducibility and support future research,
we make the source code of \mytool{} publicly available at
https://github.com/Ryan-hub-bit/numAnchor27.
The accompanying large-scale datasets and trained model
artifacts are hosted separately on
https://zenodo.org/records/19748674.

\section*{Acknowledgments}

We thank all ICSE anonymous reviewers for their valuable comments to improve this paper. 
This work is supported by NSF grants 2417055 \& 2555093 and Google Research Scholar Award.

\bibliographystyle{IEEEtran}
\bibliography{bib/ibranch,bib/path-exploration,bib/binary-code,bib/code-reuse,bib/debloating,bib/mips,bib/other,bib/jiang,bib/others,bib/gnn,bib/binary,bib/asplos,bib/addressaware}

\end{document}